\documentclass[prb,twocolumn,aps]{revtex4}
\usepackage{amsmath}
\usepackage{graphicx}
\usepackage{xcolor}
\allowdisplaybreaks{}
\def\sech{\operatorname{sech}}
\def\atan{\operatorname{atan}}
\begin{document}
\title{Fractional excitations in one-dimensional fermionic
  superfluids}

\author{Fei Ye} 

\affiliation{Department of Physics, South University of Science
  and Technology of China, Shenzhen 518055, China}

\author{P. A. Marchetti} 

\affiliation{Dipartimento di Fisica e Astronomia, INFN, I-35131 Padova, Italy}
\begin{abstract}
  We study the soliton modes carrying fractional quantum numbers in
  one-dimensional superfluids.  In the $s$-wave pairing superfluid with
  the phase of the order parameter twisted by opposite angles
  $\pm \varphi/2$ at the two ends there is an emergent complex $Z_2$
  soliton mode carrying fractional spin number $\varphi /(2\pi)$ if
  there is only one pairing branch. We demonstrate that in finite
  systems of length $L$, the spin density for one pairing branch in the
  presence of a single soliton mode consists of two terms, a localized
  spin density profile carrying fractional quantum number
  $\varphi/(2\pi)$, and a uniform background $-\varphi /(2\pi L)$. The
  latter one vanishes in the thermodynamic limit leaving a single
  soliton mode carrying fractional excitation, however it is essential
  to keep the total quantum number conserved in finite systems. This
  analysis is also applicable to other systems with fractional quantum
  numbers, thus provides a mechanism to understand the compatibility of
  the emergence of fractional charges with the integral quantization of
  charges in a finite system. For the $p$-wave pairing superfluid with
  the chemical potential interpolating between the strong and weak
  pairing phases, the $Z_2$ soliton is associated with a Majorana zero
  mode. By introducing the dimension density, we argue that the Majorana
  zero mode may be understood as an object with 1/2 dimension of the
  single particle Hilbert space. We conjecture a connection of the
  dimension density of one-dimensional solitons with the quantum
  dimension of topological excitations.
\end{abstract}

\maketitle
\section{introduction}

A fractional excitation is an emergent quasiparticle which carries only
part of the degrees of freedom of the constituent elementary particles
of the system. The best known examples in condensed matter systems
include the spin-charge separation in polyacetylene\cite{su1980}, the
quasiparticle/quasiholes in fractional quantum Hall
effect\cite{Laughlin1983}, the Majorana zero modes or Majorana fermions
in one-dimensional fermion systems\cite{Read2000,Sengupta2001,
  kitaev2001, ye2002, ye2003b}. The first theoretical model of
fractionalization was given by Jackiw and Rebbi\cite{jackiw1976}, which
is a one-dimensional model of Dirac fermion field coupled to a real Bose
field. They showed that if the Bose field configuration has a kink such
a model in the semiclassical approximation possesses a $Z_2$ soliton
mode carrying half fermion number\cite{jackiw1976, qi2008}. Later, it
has also been proved in Ref.\onlinecite{Froehlich1988} that, when the
dynamics of the Bose field is treated in a full Quantum Field Theory
framework beyond the semi-classical approximation, one can construct a
quantum kink field operator creating relativistic particles with one
half fermion number.  Furthermore the Hilbert space of states of the
model contains sectors with half-integer fermion number.

In the context of condensed matter systems, it has been found that the
organic conductor polyacetylene may be described by the electron-phonon
coupled model, where the Bose field is the optical phonon representing
the alternating displacement of ions and the corresponding soliton mode
has charge $e/2$\cite{su1980}(ignoring the fermion doubling). This model
displays strong analogies with the Jackiw-Rebbi model
\cite{Jackiw1981}. However, in the solid-state systems the basic unit of
charge is the electronic charge $e$, a natural question to ask is how to
balance the fractional charge with integral multiples of charge $e$.  In
the thermodynamic limit for the vacuum sector this problem may be solved
trivially by creating soliton and antisoliton in pairs, so that the
total charge number is still an integer. While, this argument does not
apply to a single soliton excitation appearing in the soliton sector,
which is well defined as shown in Ref.\onlinecite{Froehlich1988} in the
similar Jackiw-Rebbi model.
 
In this article we scrutinize this problem and find that in a finite
system with length $L$, when a localized fractional charge $c$ is
created, a uniform charge density $-c/L$ is left in the background which
cancels the local fractional charge after integrating over the whole
space. However for infinite systems, the thermodynamic limit
$L\rightarrow \infty$ should be taken on the correlation functions of
the local fields, and only when this limit has already been taken one
may compute the global quantities. Therefore, since the homogeneous
contribution vanishes in the thermodynamic limit, one recovers the
fractional charge of the soliton sector for infinite systems,
consistently with the results of the previously quoted references.

To investigate the fractional excitation, instead of the polyacetylene
model we consider the (quasi-)one-dimensional superfluids with two
species of fermions, which display features similar to those appearing
in Jackiw-Rebbi/polyacetylene model at the mean field level. The order
parameter plays the role of bosonic background, which can be generated
either by the emergence of degenerate ground states, leading to the
spontaneous symmetry breaking in the thermodynamic limit
\cite{Strocchi2008}, or by the proximity effect. The one-dimensional
superconductor by itself is also a source of great interest in recent
years. A reason is that it provides a candidate to study the
Fulde-Ferrell-Larkin-Ovchinnikov state \cite{fulde1964,larkin1965} of
the imbalanced superfluids with the coexistence of the superfluidity and
magnetism \cite{yang2001, orso2007, hu2007, feiguin2007, rizzi2008,
  tezuka2008, fei2009, youichi2009, Liao2010}. Another reason of
interest is that, if the pairing symmetry is $p$-wave, it may host the
Majorana zero modes \cite{Mourik2012, Das2012, Finck2013, Churchill2013,
  Deng2012, Albrecht2016} with possible non-abelian statistics useful
for the fault-tolerant quantum computation\cite{Nayak2008}.

There are some differences between the fermionic superfluids and the
polyacetylene. In the polyacetylene, the bosonic field (dimerization
parameter) is real, i.e., either positive or negative, and the
corresponding soliton belongs to the $Z_2$ class, while in the $s$-wave
superconductor, the order parameter $\Delta(x)$ is complex so that it is
possible to generate a soliton with an arbitrary phase difference
$\varphi$ between the two ends of the superconducting nanowire, which we
call \emph{complex $Z_2$ soliton}(see Sec.~\ref{sec:hamiltonian}).  The
fractional quantum number carried by this soliton mode is
$\varphi/(2\pi)$\cite{goldstone1981, mackenzie1984}. Furthermore, in
polyacetylene, both the charge and spin are conserved leading to the
spin-charge separated excitations, while in one-dimensional
superconductors, the charge conservation is broken in the mean-field
treatment, and only the spin number is conserved, therefore the
fractionalized quantum number is actually the electronic spin.  Note
that the total spin number in the $s$-wave superconductor is invariant
in the process of twisting the phase difference $\varphi$ continuously
in a finite system. Since the system is uniform with zero spin when
$\varphi=0$, then a question arises: how a single soliton mode with
fractional quantum number $\varphi/(2\pi)$ can emerge while still
keeping the total quantum number as an invariant integer?

For the $p$-wave pairing superconductors, even the spin number is not
conserved anymore, and only the fermion number parity does. The
corresponding $Z_2$ soliton is of Majorana type. In the following
sections, we provide a systematic investigation of these questions.

This paper is organized as follows. In section \ref{sec:soliton-mode-S},
we discuss the soliton mode in the one-dimensional $s$-wave
superconductor.  A brief introduction to the Hamiltonian and the
notations is given in section \ref{sec:hamiltonian}, and we consider the
single complex $Z_2$ soliton excitation, discussing its energy and the
effect of finite momentum cutoff in \ref{sec:wavef-phase-shift} and
\ref{sec:soliton-energy}, and the spin density distribution in
\ref{sec:fract-spin-distr}. In section \ref{sec:majorana-zero-P}, we
consider the Majorana zero mode in $p$-wave pairing superconductor. The
section \ref{sec:conclusion} is a brief summary of the conclusions.

\section{Soliton mode in the one dimensional $s$-wave superconductor}
\label{sec:soliton-mode-S}
\subsection{Hamiltonian}
\label{sec:hamiltonian}
We consider the one-dimensional electron gas with attractive
interaction, whose low energy behavior is controlled by the
quasiparticles near the two Fermi points with linear spectra. Thus, it
can be effectively described in terms of four chiral Fermi fields, the
right movers $\hat{R}_{\sigma}(x)$ and the left movers
$\hat{L}_{\sigma}$ with spin indexes $\sigma=\uparrow,\downarrow$. In
the $s$-wave BCS theory, the pairing takes place between
$\hat{R}_{\sigma}(x)$ and $\hat{L}_{-\sigma}(x)$. The corresponding mean
field Hamiltonian is given by
\begin{eqnarray}
  \label{eq:1}
  \hat{H}=\int dx [\hat{\mathcal{H}}_1(x) + \hat{\mathcal{H}}_2(x)]  + \int dx \frac{|\Delta(x)|^2}{g} 
\end{eqnarray}
where $g(>0)$ is the attractive interaction. The Hamiltonian density
$\hat{\mathcal{H}}_1$ and $\hat{\mathcal{H}}_2$ read

\begin{align}
  \label{eq:2}
  &\hat{\mathcal{H}}_1(x)=-i\hat{R}^{\dagger}_{\uparrow}\partial_x
  \hat{R}_{\uparrow}
  +i\hat{L}^{\dagger}_{\downarrow}\partial_x \hat{L}_{\downarrow}  \nonumber\\
  &
  \hspace{2.0cm}+\Delta(x) \hat{R}_{\uparrow}^{\dagger} \hat{L}^{\dagger}_{\downarrow}
  +\Delta^{*}(x) \hat{L}_{\downarrow}\hat{R}_{\uparrow}
  \\
\label{eq:3}
  &\hat{\mathcal{H}}_2(x)=i\hat{L}^{\dagger}_{\uparrow}\partial_x
  \hat{L}_{\uparrow}
  -i\hat{R}^{\dagger}_{\downarrow}\partial_x \hat{R}_{\downarrow} \nonumber\\
  &\hspace{2.0cm} +
  \Delta(x) \hat{L}_{\uparrow}^{\dagger} \hat{R}^{\dagger}_{\downarrow}
  +\Delta^{*}(x) \hat{R}_{\downarrow}\hat{L}_{\uparrow}
\end{align}
where both the Planck constant $\hbar$ and the Fermi velocity
$v_F$ are set to 1. The order parameter is determined self-consistently by
the variational principle
\begin{eqnarray}
  \label{eq:4}
  \Delta(x) = -g[\langle \hat{L}_{\downarrow}\hat{R}_{\uparrow}
  \rangle +\langle \hat{R}_{\downarrow}\hat{L}_{\uparrow}\rangle].
\end{eqnarray} 
The Eqs. \eqref{eq:1}-\eqref{eq:4} are invariant under the $Z_2$
symmetry:
\begin{eqnarray}
  \label{eq:5}
  \Delta(x) \rightarrow \Delta^{*}(x) , \; L^\dagger_\downarrow
  \leftrightarrow R_\uparrow, \; L^\dagger_\uparrow \leftrightarrow R_\downarrow.
\end{eqnarray}
This symmetry guarantees that the groundstates with $\Delta$ and
$\Delta^*$ are degenerate in energy, but a generic phase difference in
the choice of $\Delta$ not respecting the $Z_2$ symmetry yields a
different groundstate energy. Therefore a kink can only interpolate
between $\Delta$ and $\Delta^*$, not between arbitrary phases for
$\Delta$. In spite of the appearance of complex phases, the kink is still
related to a $Z_2$ symmetry. We then call such a soliton 
\emph{complex $Z_2$ soliton}.

Given an order parameter $\Delta(x)$, one may write the Hamiltonian
density $\hat{\mathcal{H}}_1$ in diagonalized form
\begin{eqnarray}
  \label{eq:6}
  \int  dx\hat{\mathcal{H}}_1(x) = \sum_n \frac{\epsilon_{n}}{2}[
  \hat{d}_{1n}^{\dagger}\hat{d}_{1n} -
  \hat{d}_{1n}\hat{d}_{1n}^{\dagger}]
\end{eqnarray}
via the substitution
$\hat{R}_{\uparrow} (x) = \sum_n \hat{d}_{1n} u_n(x) $ and
$\hat{L}^{\dagger}_{\downarrow} (x) = \sum_n \hat{d}_{1n} v_n(x)$. The
spinor $(u_n(x),v_n(x))^{t}\equiv\phi_n(x)$ satisfies the following
differential equation,
\begin{eqnarray}
  \label{eq:7}
  [-i\partial_x\sigma_3+\Delta_1\sigma_1+\Delta_2\sigma_2]\phi_n(x) = \epsilon_n\phi_n(x),
\end{eqnarray}
$\Delta_{1,2}(x)$ being the real and imaginary part of
$\Delta(x)$, respectively. Using Eq.~\eqref{eq:7}, it is easy to
check that $\sigma_2\phi^{*}_n(x)$ satisfies
\begin{eqnarray}
  \label{eq:8}
  [i\partial_{x}\sigma_3 +\Delta_1\sigma_1
  +\Delta_2\sigma_2]\sigma_2\phi^{*}_n(x)
  = -\epsilon_n\sigma_2\phi^{*}_n(x),
\end{eqnarray}
which is exactly the equation for single-particle eigenfunction of
$\hat{\mathcal{H}}_2$. Thus, we substitute
$\hat{L}_{\uparrow}(x) = -\sum_n \hat{d}_{2n} v^{*}_n(x) $ and
$ \hat{R}^{\dagger}_{\downarrow}(x) = \sum_n \hat{d}_{2n} u^{*}_n(x) $
into $\hat{\mathcal{H}}_2$ and obtain the following diagonalized form
\begin{eqnarray}
  \label{eq:9}
  \int dx \hat{\mathcal{H}}_2(x)  = -\sum_n \frac{\epsilon_n}{2}
  [\hat{d}^{\dagger}_{2n} \hat{d}_{2n} -
  \hat{d}_{2n}\hat{d}_{2n}^{\dagger} ]\;. 
\end{eqnarray}

To create a soliton in a finite system of length $L$, we impose a
twisted boundary condition on the order parameter $\Delta(x)$,
\begin{eqnarray}
  \label{eq:10}
  \Delta(\pm \frac{L}{2}) = \Delta_0 e^{\pm i\varphi/2},
\end{eqnarray}
and $\partial_x\Delta(\pm \frac{L}{2})$ should tend to zero in the
thermodynamic limit. Since the order parameter is determined by
Eq.~\eqref{eq:4}, the twist boundary condition Eq. \eqref{eq:10} can
also be implemented on the wavefunctions $\phi_n$:
\begin{eqnarray}
  \label{eq:11}
  \phi_n(-L/2)= 
  e^{-i\sigma_3\varphi/2}\phi_n(L/2)\;.
\end{eqnarray}
It is straightforward to verify the solutions of Eq.~\eqref{eq:8},
$\sigma_2\phi^{*}_n(x)$, also satisfy this boundary condition.

If $\varphi = 0$, the superfluid is homogeneous with a real positive
order parameter $\Delta_0$ determined by
\begin{eqnarray}
  \label{eq:12}
  1 =  \frac{g}{\pi}
  \int_0^{\Lambda} \frac{dk}{\omega(k)},
\end{eqnarray}
where $\omega(k) = \sqrt{k^2+\Delta_0^2}$ and $\Lambda$ is the momentum
cutoff. The groundstate energy is given by
\begin{eqnarray}
  \label{eq:13}
  E_g= - \sum_n {|\epsilon_n|},
\end{eqnarray}
which is not a simple summation over all negative eigenvalues, but also
includes the energy of fermion vacuum disturbed by the inhomogeneous
$\Delta(x)$ if $\varphi \neq 0$.

\subsection{Wavefunctions and phase shift}
\label{sec:wavef-phase-shift}
The mean-field Hamiltonian Eq.~\eqref{eq:1} can be solved with the
inverse scattering method. Some relevant results are given here, and one
can refer to Refs.~\onlinecite{frolov1972, dashen1975, shei1976,
  grosse1987} for more details on the inverse scattering method and its
application to one-dimensional Dirac models.

For convenience we first permute the Pauli matrices:
$\sigma_1\rightarrow\sigma_{3}$, $\sigma_2\rightarrow\sigma_{1}$ and
$\sigma_3\rightarrow\sigma_2$, so that the Dirac Hamiltonians in
Eqs.~\eqref{eq:7} and \eqref{eq:8}, denoted by $\mathbf{H}_1$ and
$\mathbf{H}_2$, respectively, become purely \emph{real}:
\begin{align}
  \label{eq:14}
  &\mathbf{H}_1=
  \begin{pmatrix}
    \Delta_1 & -\partial_x +\Delta_2 \\ \partial_x+\Delta_2 & -\Delta_1
  \end{pmatrix}, \\
\label{eq:15}
  & \mathbf{H}_2=\begin{pmatrix}
    \Delta_1 & \partial_x +\Delta_2 \\ -\partial_x+\Delta_2 & -\Delta_1
  \end{pmatrix}.
\end{align}
Then, it is obvious that the eigenfunctions of $\mathbf{H}_1$ (or
$\mathbf{H}_2$) appear in conjugate pairs with two-fold degeneracy. We
note that such a permutation is actually equivalent to a unitary
transformation $U=e^{i\pi/(3 \sqrt{3})(\sigma_1+\sigma_2+\sigma_3)}$.

Since $\mathbf{H}_{2}=-\sigma_2 \mathbf{H}_1\sigma_2$, and we can focus
on the $\mathbf{H}_1$-branch only.  The eigenfunctions $\psi_{n}$ of
$\mathbf{H}_1$ are related to the original ones $\phi_n$ by the
unitary transformation $\psi_n=U\phi_n$, and, according to
Eq.~\eqref{eq:11}, the $\psi_n$'s satisfy the boundary condition
\begin{eqnarray}
  \label{eq:16}
  \psi_n(-L/2) = e^{i\sigma_2\varphi/2}
  \psi_n(L/2). 
\end{eqnarray}
Mathematically, it guarantees the Dirac operator to be Hermitean. For
$\varphi=0$, the ground state is a simple BCS state with a constant
order parameter $\Delta_0$ which is assumed to be positive, and there is
no midgap state. If $\varphi \ne0$, the order parameter $\Delta(x)$,
which minimizes the free energy, is reflectionless, and it simplest form
could be
$\Delta(x)=\epsilon-i\kappa\tanh(\kappa
x)$\cite{shei1976,grosse1987}. The parameters $\epsilon$ and $\kappa$
are determined by the boundary condition Eq.~\eqref{eq:10} as follows:
\begin{eqnarray}
  \label{eq:17}
  \epsilon=\Delta_0\cos \frac{\varphi}{2}, \hspace{0.5cm}
  \kappa=\Delta_0\sin \frac{\varphi}{2}. 
\end{eqnarray}
The phase angle $\varphi$ is restricted in the range $0\le\varphi<2\pi$,
therefore $\kappa\ge0$. Thus, $\mathbf{H}_1$ can be rewritten as
\begin{eqnarray}
  \label{eq:18}
  \mathbf{H}_1 = \begin{pmatrix}
    \epsilon & -\partial_x+\kappa\tanh(\kappa x) \\ 
    \partial_x+\kappa\tanh(\kappa x) & -\epsilon
  \end{pmatrix}.
\end{eqnarray}
Its spectrum consists of three parts, the negative scattering continuum
with energy $\omega\le -\Delta_0$, a single midgap state with energy
$\epsilon$, and the positive scattering continuum with energy
$\omega\ge\Delta_0$. Similar results were also given in
Refs.~\onlinecite{shei1976, grosse1987} in a different form.  

In the uniform ($\varphi=0$) and the Jackiw-Rebbi soliton ($\varphi=\pi$)
phases, the system possesses a particle-hole symmetry in the
thermodynamic limit. The charge conjugation may be defined as
$\psi\rightarrow\sigma_1\psi$ for $\varphi=0$, and
$\psi\rightarrow\sigma_3\psi$ for $\varphi=\pi$. For a general $\varphi$
with $\epsilon\ne0$, one can not find such a particle-hole
transformation.

\begin{figure}[htbp]
 \centerline{\includegraphics[width=6cm]{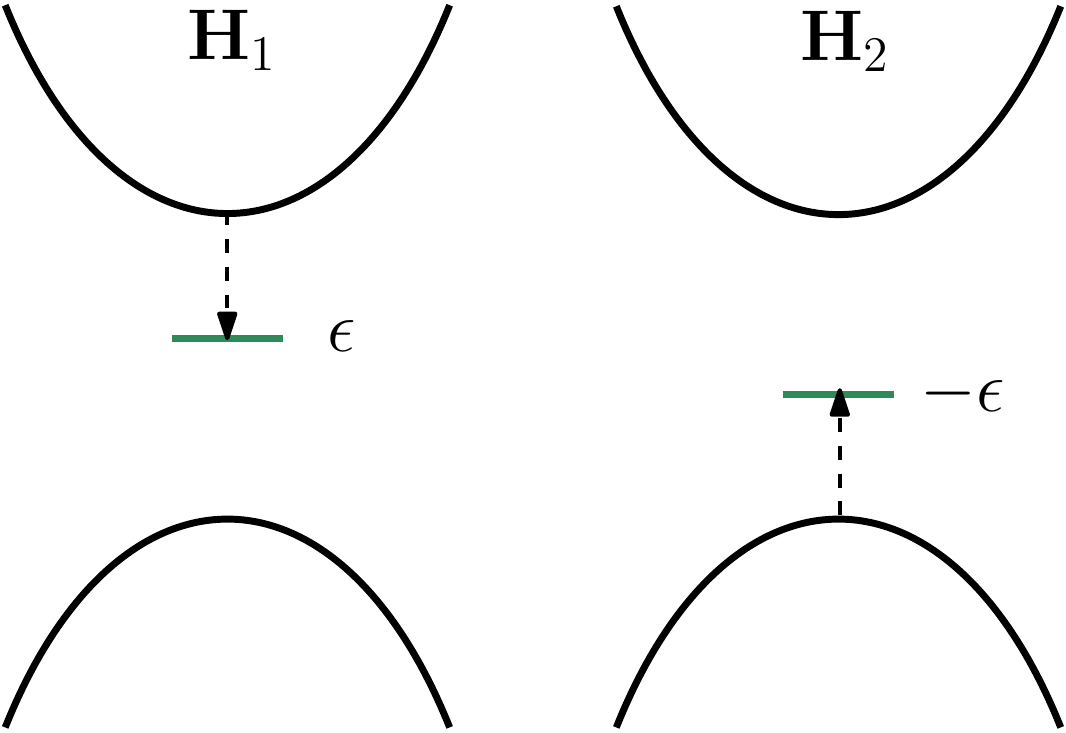}}
  \caption[]{\label{fig:gapstate} Illustration of evolution of midgap
    states of $\mathbf{H}_1$ and $\mathbf{H}_2$ branches as the twisted
    angle is increasing.}
\end{figure}

The complete set of eigenpairs of the Hamiltonian of Eq.~\eqref{eq:18}
is listed below. The midgap state has a localized wavefunction given in
the thermodynamic limit by
\begin{eqnarray}
  \label{eq:19}
  \psi^B(x,\kappa) = \sqrt{\frac{\kappa}{2}} \sech(\kappa x) \begin{pmatrix}
    1\\ 0
  \end{pmatrix}
\end{eqnarray}
with energy $\epsilon$. The characteristic width of the soliton is given
by $1/\kappa$. Note that Eq.~\eqref{eq:19} satisfies the boundary
condition Eq.~\eqref{eq:16} approximately with an exponentionally small
error $e^{-\kappa L}$ which does not produce anything significant in our
calculations when $L\gg \kappa^{-1}$. As the twisted angle is increasing
from $0$, this midgap state of $\mathbf{H}_{1}$ is lowered down from the
bottom of conduction band, meanwhile the midgap state of $\mathbf{H}_2$
with energy $-\epsilon$ is lifted from the top of valence band as
illustrated in Fig. \ref{fig:gapstate}.

The scattering wavefunctions are given by
\begin{eqnarray}
  \label{eq:20}
  \psi^S_{\pm}(x,k)= N_{\pm}^S e^{ikx} \begin{pmatrix}
    \kappa\tanh(\kappa x)-ik \\ \pm \omega-\epsilon
  \end{pmatrix},
\end{eqnarray}
with eigenvalues $\pm\omega\equiv \pm\sqrt{k^2+\Delta_0^2}$ and
normalization constant
$N^S_{\pm} = [2L\omega(\omega\mp\epsilon)-2\kappa\tanh(\kappa
L/2)]^{-1/2}$.

To determine the scattering phase shift from the kink of $\Delta(x)$,
let's first check the asymptotic behaviors of $\mathbf{H}_1$ as
 $x\rightarrow\pm\infty$
\begin{align}
  \label{eq:21}
  \mathbf{H}_{1,\pm\infty} =&
                              -i\partial_x\sigma_2\pm\kappa\sigma_1+\epsilon\sigma_3 
\end{align}
which differ from each other by a SU(2) rotation
$e^{i\sigma_2\varphi/2}$:
$e^{i\sigma_2\varphi/2}\mathbf{H}_{1,\infty}e^{-i\sigma_2\varphi/2}=
\mathbf{H}_{1,-\infty}$. For convenience, we denote the asymptotic
wavefunction by $\psi^S_{\pm\infty}(x)$ at $\pm\infty$.  Note that
$e^{i\sigma_2\varphi/2}\psi^S_{\infty}(x)$ is an eigenstate of
$\mathbf{H}_{1,-\infty}$, thus it only differs from
$\psi^S_{-\infty}(x)$ by a phase shift $\delta(k)$, namely,
\begin{eqnarray}
  \label{eq:23}
 e^{i \sigma_2 \varphi/2}\psi^S_{\infty}(x,k)
  =e^{i\delta(k)} \psi^{S}_{-\infty}(x,k).
\end{eqnarray}
Such a definition of phase shift has also been adopted for the $Z_2$
kink in the polyacetylene in Ref.~\onlinecite{kivelson1982} together
with its physical consequences.

Using the scattering wavefunctions Eq.~\eqref{eq:20}, we obtain the
phase shift as follows:
\begin{align}
  \label{eq:24}
  \delta^{(+)}(k) =& \pi-\atan \frac{k}{\kappa} -\atan \frac{\epsilon
                     k}{\omega\kappa}, \nonumber\\
&\hspace{2cm}
  \text{ for the conduction band, } \nonumber\\
\delta^{(-)}(k)= & -\atan \frac{k}{\kappa} +\atan \frac{\epsilon
                   k}{\omega\kappa}, \nonumber\\
&\hspace{2cm}
  \text{ for the valence band,} 
\end{align}
as given previously in Ref.~\onlinecite{shei1976}. The phase
shift Eq.~\eqref{eq:24} is useful for calculating the soliton energy
and the fractional spin density distribution.

For $\mathbf{H}_2$, since $\mathbf{H}_2=-\sigma_2\mathbf{H}_1\sigma_2$,
its eigenfunctions are simply $\sigma_2\psi^B$ and $\sigma_2\psi^S$ with
opposite eigenvalues of those of $\mathbf{H}_1$.

\subsection{Soliton energy}
\label{sec:soliton-energy}

The groundstate energy of the uniform BCS state is divergent as the
momentum cutoff $\Lambda\rightarrow\infty$, and this divergence persists
in the presence of a soliton. However the difference of these energies
is finite and may be defined as the excitation energy of the soliton
mode. There are two contributions to the soliton energy, one is the
readjustment of the fermionic spectra in the presence of the soliton,
the other is the modification of the condensate energy.

To calculate the excitation energy of the soliton we follow the method
given in Refs.~\onlinecite{dashen1975} and \onlinecite{shei1976}. The
basic idea is to discretize the pseudo-momentum $k$ by putting the
system into a box with length $L$ which is required to be much larger
than the soliton size, i.e., $L\gg\kappa^{-1}$, and imposing the
boundary condition Eq.~\eqref{eq:16}. Thus, it is possible to make a
one-to-one correspondence between the spectra of the uniform state and
those of the soliton state. In the homogeneous state without the phase
shift, the quantized momentum takes the values of $k_n = 2n\pi/L$. When
the twisted angle $\varphi$ is turned on adiabatically starting from
zero, the pseudo-momentum $\bar{k}_n$ is shifted from $k_n=2n\pi/L$ and
satisfies
\begin{eqnarray}
  \label{eq:25}
  \bar{k}^{(\pm)}_nL + \delta^{(\pm)}(\bar{k}^{(\pm)}_n) = k_nL,
\end{eqnarray}
where the phase shifts $\delta^{(\pm)}(k)$ are given in
Eq.~\eqref{eq:24}. Therefore, the energy for each pseudo-momentum in the
presence of the soliton is shifted from that in the uniform state.  It
is then straightforward to calculate the soliton energy by collecting
the energy shifts for all the pseudo-momenta within the momentum
cutoff; it is given by
\begin{eqnarray}
\label{eq:26}
&&E_s(\Lambda) = 
\frac{2\omega_c}{\pi}\atan \frac{\kappa}{\Lambda}+
\frac{2\epsilon}{\pi} \atan \frac{\Lambda\epsilon}{\kappa\omega_c} -
|\epsilon| \nonumber\\
&&\lim_{\Lambda\rightarrow\infty} E_s(\Lambda) =
\frac{2\kappa}{\pi} + \frac{2\epsilon}{\pi} \atan
\frac{\epsilon}{\kappa} - | \epsilon|,
\end{eqnarray}
where $\omega_c=\sqrt{\Lambda^2+\Delta_0^2}$. If $\varphi=\pi$, we
obtain the well known result $E_s(\infty)=2\Delta_0/\pi$ for the
Jackiw-Rebbi soliton.

\subsection{Fractional spin excitations}
\label{sec:fract-spin-distr}
We introduce two local spin density operators as follows:
\begin{eqnarray}
  \label{eq:27}
  \hat{s}_1(x) = \hat{R}^{\dagger}_{\uparrow} \hat{R}_{\uparrow} -
  \hat{L}^{\dagger}_{\downarrow} \hat{L}_{\downarrow}, \hspace{0.2cm}
  \hat{s}_2(x) = \hat{L}^{\dagger}_{\uparrow} \hat{L}_{\uparrow} -
  \hat{R}^{\dagger}_{\downarrow} \hat{R}_{\downarrow},
\end{eqnarray}
which corresponds to the two pairing branches, respectively.  The
expectation values of $\hat{s}_{1,2}$ denoted by $s_{1,2}(x)$ can be
expanded in terms of the eigenmodes,
\begin{align}
\label{eq:28}
s_1(x)=&
\sum_n \langle \hat{d}^{\dagger}_{1n} \hat{d}_{1n} \rangle
|\phi_n(x)|^2 - \frac{1}{2}
\sum_n |\phi_n(x)|^2, \\
\label{eq:29}
s_2(x) =&\sum_n \langle \hat{d}^{\dagger}_{2n} \hat{d}_{2n} \rangle
|\phi_n(x)|^2 - \frac{1}{2}
\sum_n |\phi_n(x)|^2 .
\end{align}

In Eqs.~\eqref{eq:28} and \eqref{eq:29}, the $\phi_n$'s can be replaced
by the orthonormal wavefunctions $\psi_n$'s since $\phi_n$'s are related
to $\psi_n$'s by a unitary transformation.  The summations in the
equations above would diverge without cutoff, hence we regulate them by
introducing an UV momentum cutoff $\Lambda$ chosen in such a way that in
the homogeneous state there are $2N+1$ energy eigenstates for both the
valence and the conduction bands.  Thus in the homogeneous state, the
second term in Eqs. ~\eqref{eq:28} and \eqref{eq:29} is equal to
$-(2N+1)/L$. In the presence of the soliton, this term still has a
leading contribution $-(2N+1)/L$ in an expansion in the inverse powers
of $L$ as shown in the appendix \ref{sec:deriv-spin-dens}, as long as we
keep track of each quantum state in the process of increasing $\varphi$.

We now calculate $s_1(x)$ by assuming that all the scattering states in
the valence band are occupied. The midgap state of the
$\hat{\mathcal{H}}_1$-branch is left empty since it is connected to the
bottom state in the conduction band and remains empty in the adiabatic
process of increasing the twisted phase angle $\varphi$. The spin
density $s_1(x)$ is then given by
\begin{align}
\label{eq:30}
s_1(x) =& \sum_{n}|\psi^S_{-}(x,\bar{k}^{(-)}_n)|^2
-\frac{2N+1}{L}.
\end{align}

Clearly, the total spin number $\int_{-L/2}^{L/2} dx s_1(x) = 0$, since
the eigenstates are normalized. This agrees with \emph{the conservation
  law of the total spin} in the adiabatic process of tuning the phase
$\varphi$ of $\Delta(x)$ away from the uniform BCS state in a finite
system, since the mean-field Hamiltonian always commute with the total
spin operator irrespective of the specific form of $\Delta(x)$. Any kind
of excitations upon this groundstate creates an integer number of
spins. This results in a puzzle: can we get a fractional excitation in a
finite system, while still respecting the conservation law of fermion
number? The answer is yes, as we show below by calculating the local
spin density $s_1(x)$ the exact orthonormal wavefunctions followed by an
expansion with respect to $1/L$ for large enough $L$.

Substituting Eq.~\eqref{eq:20} into Eq.~\eqref{eq:30} and considering
the allowed $\bar{k}_n$ given by Eq.~\eqref{eq:25}, one finds that, up
to higher order terms, the spin density $s_1(x)$ in fact consists of two
parts, a uniform term $s_{1u}$ and a localized term $s_{1l}$:
\begin{align}
\label{eq:31}
& s_1(x) = s_{1u} + s_{1l}(x)\;, \nonumber\\
& s_{1u} = \frac{c}{L}\;,  \nonumber\\
& s_{1l}(x) = - \frac{c\kappa}{2} \sech^2(\kappa x) \;,
\end{align}
where $c\equiv\varphi/(2\pi)$. The detailed calculation is given in the
appendix \ref{sec:deriv-spin-dens}.

Eq.~\eqref{eq:31} implies that, if one can measure the spin density with
spatial resolution high enough, a localized spin distribution will be
found around the origin($x=0$) carrying a fractional spin number $-c$
with density profile described by $s_{1l}(x)$, while the uniform one
$s_{1u}$ can be ignored in this local measurement, being $O(1/L)$. Thus,
it is legitimate to identify the fractional excitation as the localized
density profile $s_{1l}(x)$ in a finite system. Since the thermodynamic
limit is taken on local correlators for the infinite volume, the uniform
term $s_{1u} $ completely disappears and we get a true fractional
charge, in agreement with the results of Refs.\onlinecite{jackiw1976}
and \onlinecite{Froehlich1988}. The contribution of $s_{1u}$, which
compenstates that of the localized term $s_{1l}$ after integration,
however, can not be ignored in the discussion of the conservation of the
total quantum number in finite systems.

In Eq.~\eqref{eq:30}, $s_1(x)$ is calculated with the midgap state
unoccupied and shows a localized spin distribution with a fractional
quantum number $-c$.  If the midgap state is filled, the localized spin
density becomes $(1-c)\kappa\sech^2(\kappa x)/2$ with fractional quantum
number $1-c$. If $c=1/2$, one recovers the well known result for the
Jackiw-Rebbi soliton: the soliton mode carries $\pm 1/2$ charge
depending on whether the zero energy state is filled or not. For general
$c$, the particle-hole symmetry is broken. 

Usually, one uses the arguments with fermion doubling or multiple
soliton (antisoliton) excitations to resolve the issue concerning the
compatibility of the emergence of the fractional fermion number with the
conservation of the integer number of the elementary constituent
fermions. Here, we provide a new and more fundamental perspective to
this issue that a single localized soliton mode carrying a fractional
quantum number can exist independently (in the sense specified above) in
a one-dimensional finite system compenstated by an opposite fractional
charge distributed uniformly in the background.

Similarly one can compute the spin density $s_2(x)$ of the other pairing
branch.  Note that the wavefunctions of the valence band of
$\hat{\mathcal{H}}_2$ are $\sigma_2\psi^{(+)}(x,\bar{k}^{(+)}_n)$, and
the midgap state is $\sigma_2\psi^B(x)$, which evolves from the state on
the top of the valence band and is assumed to be occupied as we tune the
phase $\varphi$ adiabatically. Then, we obtain
\begin{eqnarray}
\label{eq:32}
s_2(x) = \sum_{n}|\psi^S_{+}(x,\bar{k}_n)|^2
+ |\psi^B(x)|^2 - \frac{2N+1}{L}.
\end{eqnarray}
As shown in the appendix ~\ref{sec:deriv-spin-dens}, to the leading
order, $s_2(x)$ can be recast as
\begin{align}
\label{eq:33}
&s_2(x) = s_{2u} + s_{2l}(x), \nonumber\\
& s_{2u}= - \frac{c}{L}, \nonumber\\
&s_{2l} = \frac{\kappa c}{2}\sech^2(\kappa x).
\end{align}
Again, we observe a localized fractional excitation $s_{2l}$ and a
uniform background $s_{2u}$. They cancel with each other after
integration respecting the spin conservation law in a finite system.

Although there are fractional excitations for each branch individually,
their sum is trivial $s_1(x)+s_2(x)=0$ consistent with the result for
the uniform BCS state.  In fact, there are four cases with different
spin number depending on the configurations of the midgap states as
follows: spin 1 for both filled, spin 0 for half filling with two cases,
and spin -1 for both empty, all of which carry an integer spin number
instead of a fractional one. Thus, the fermion doubling dispels the
fractional excitation. To get the fractional spin excitations, one needs
to avoid the fermion doubling. Note that the edge mode of a quantized
spin Hall insulator consists of only one branch of Dirac
fermion\cite{Kane2005b, Bernevig2006}, if it is coupled to a
superconductor appropriately, the proximity effect may induce a
one-dimensional superconductor without fermion doubling which may be a
possible candidate to observe the fractional spin excitation.

\section{Majorana zero mode in $p$ wave superfluid}
\label{sec:majorana-zero-P}
In this section, we consider the soliton mode in one-dimensional
$p$-wave superconductors with only one species of fermions with the
following minimal Hamiltonian\cite{Read2000,Sengupta2001}
\begin{align}
\label{eq:34}
  \hat{H}=&-\int dx \mu(x)\hat{c}^{\dagger}(x)\hat{c}(x) \nonumber\\
     &-\frac{1}{2}\int dx \left[\hat{c}^{\dagger}(x)i\partial_x\hat{c}^{\dagger}(x)+\hat{c}(x)i\partial_x\hat{c}(x) \right]
\end{align}
where $\hat{c}(x)$ and $\hat{c}^{\dagger}(x)$ are the spinless
annhilation and creation operators. The chemical potential
$\mu(x)=\mu_0\tanh(\mu_0x)$ with $\mu_0>0$, interpolates between the
weak and strong pairing phases. This Hamiltonian resembles the
Jackiw-Rebbi model, and a $Z_2$ soliton mode occurs which turns out to
be a Majorana zero mode.

The Hamiltonian Eq.~\eqref{eq:34} can be diagonalized as
\begin{align}
\label{eq:35}
\hat{H}=\sum_k \omega_k \hat{d}_k^{\dagger}\hat{d}_k+\text{const.}
\end{align}
where $\omega_k=\sqrt{k^2+\mu_0^2}$ for the scattering state with
$k\ne0$, and $\omega_0=0$ for the zero mode.  The quasi-particle
operator $\hat{d}_k$ is a linear combination of $\hat{c}(x)$ and
$\hat{c}^{\dagger}(x)$ given by
\begin{align}
\label{eq:36}
\hat{d}_k = \int dx[ u^{*}_k(x) \hat{c}(x) + v_k(x) \hat{c}^{\dagger}(x)].
\end{align}
The wavefunction $\phi(x,k)\equiv[u^{*}_k(x),v_k(x)]^t$ satisfies the
following equation
\begin{align}
  \label{eq:37}
\mathcal{H}\phi(x,k)\equiv [ -\mu(x)\sigma_3 + i\partial_x\sigma_1]\phi(x,k)=\omega_k\phi(x,k).
\end{align}
Note that since $\sigma_1\mathcal{H}^{*}\sigma_1=-\mathcal{H}$, the
state $\sigma_1\phi^{*}(x,k)$ has a negative energy $-\omega_k$ and the
corresponding annihilation operator is simply the Hermitean conjugate
of $\hat{d}_k$. Therefore, one need only take the positive energy states
into account.

In a finite system of length $L$ the energy eigenfunctions consist of a
localized zero mode, given, up to a rest exponentially small in $L$, by
\begin{align}
\label{eq:38}
 \phi^B(x) =& \frac{e^{-i\pi/4}\sqrt{\mu_0}}{2}\sech(\mu_0x) \begin{pmatrix}
 1 \\ i
\end{pmatrix},
\end{align}
and of the scattering wavefunctions
\begin{align}
\label{eq:39}
  \phi^S(x,k) = \frac{1}{N_k}e^{-i\pi/4}e^{ikx}
  \begin{pmatrix}
 \mu(x) - ik-\omega \\ i\mu(x) + k + i\omega
\end{pmatrix},
\end{align}
with the normalization constant $N_k=2\sqrt{L\omega_k^2-\mu(L/2)}$.  The
global phase factor $e^{-i\pi/4}$ is taken for
convenience. Eq.~\eqref{eq:38} shows that $u_0(x)=v_0(x)$, which
indicates that $\hat{d}_0$ is actually a \textit{Majorana} zero mode,
while $\hat{d}_k$ is a normal Fermi operator for the scattering solution
with $k\ne0$.

Since the chemical potential has opposite signs at the two ends, a
particle excitation at one end could be a hole excitation at the other
end. This allows us to choose the boundary condition for a finite system
as
\begin{align}
\label{eq:40}
\sigma_1\phi^S(L/2,k)= \phi^S(-L/2,k).
\end{align}
Note that the inverse transformation of Eq.~\eqref{eq:36} gives rise to the
spinless fermion operator in terms of $\hat{d}_k$ as
\begin{align}
\label{eq:41}
  \hat{c}(x) =& u_0(x) \hat{d}_0 + \sum_{\bar{k}}u_{\bar{k}}(x)\hat{d}_{\bar{k}} + v_{\bar{k}}(x) \hat{d}^{\dagger}_{\bar{k}}.
\end{align}
As long as $L\gg \mu_0^{-1}$, $u_0(\pm L/2)\rightarrow 0$, and the
boundary condition Eq.~\eqref{eq:40} is equivalent to
$c(L/2)=c^{\dagger}(-L/2)$.

As in the $s$-wave pairing case, one can not define the phase shift
directly, because the asymptotic Hamiltonian as $x\rightarrow\pm\infty$
are not equal, instead they are related to each other by
$\sigma_1\mathcal{H}_{-\infty}\sigma_1=\mathcal{H}_{\infty}$. Therefore,
the phase shift can be defined as
$e^{i\delta(k)}\phi^S_{-\infty}(x,k)=\sigma_1\phi^S_{\infty}(x,k)$,
where $\phi_{\pm\infty}^S(x,k)$ are the asymptotic wavefunction as
$x\rightarrow\pm\infty$, and it can be calculated as
\begin{align}
\label{eq:42}
\delta(k) = -\pi/2-\atan(k/\mu_0).
\end{align}
For a finite system, the allowed pseudo-momentum is determined by the
boundary condtion Eq.~\eqref{eq:40} and the phase shift
Eq.~\eqref{eq:42}.

Since in the $p$-wave pairing phase, the global U(1) symmetry is broken
down to $Z_2$, the fermion number is not a conserved quantity anymore,
but its parity is still conserved. On the other hand, the dimension of
the Hilbert space should keep the same, no matter how the annihilation
and creation operators are mixed. It also does not change with the
chemical potential as long as we make a one-to-one correspondence of the
allowed pseudo-momenta for different $\mu_0$ and impose the same
boundary condition Eq.~\eqref{eq:40}.  Therefore, we suggest that the
fractionalized quantum number of the soliton in the $p$-wave pairing
superconductor is the the ``dimension'' of its \textit{single particle}
Hilbert space, defined via a suitably renormalized the trace of the
identity.

For this purpose, with an UV momentum cutoff $\Lambda$ chosen as before,
we introduce the concept of the \textit{dimension density $D(x)$ of the
  single particle Hilbert space} as the sum over the modulus square of
all the energy eigenfunctions up to the cutoff, i.e. the diagonal of the
resolution of the identity for the energy in the $x$-representation. It
follows that the integral of $D(x)$ is the total dimension of the
Hilbert space, or, equivalently, the number of the allowed
pseudo-momenta plus the number of localized zero modes, denoted by $N_D$
and it is invariant against adiabatic changes of the chemical
potential. In our model with the soliton mode
\begin{align}
\label{eq:44}
  D(x) =|\phi^B(x)|^2 +\sum_{\bar{k}} |\phi^S(x,\bar{k})|^2.
\end{align}
Its integral diverges in the limit of infinite cutoff $\Lambda$, but the
difference with respect to the same quantity in the absence of the $Z_2$
soliton is finite and independent of the cutoff.

For a normal system with fermion number conserved $D(x)$ is uniform and
equal to $N_D/L$, since after all $D(x)$ is just the fermion density
with all the single particles states occupied.  However, $D(x)$ is not
uniform in the present system with a localized Majorana zero mode (in a
sense only one half of the Jackiw-Rebbi model).  Following a procedure
similar to that given in the appendix, we find in presence of the
soliton
\begin{align}
\label{eq:45}
  D(x)-\frac{N_D}{L} = \frac{1}{2}|\phi_B(x)|^2-\frac{1}{2L} + o(L^{-2}).
\end{align}
We observe that there is a localized term which carries $1/2$ factor
indicating 1/2 dimension of the \textit{single particle} Hilbert
space. Once integrated over the whole space, it can be cancelled by the
next uniform term in the r.h.s. of Eq.~\eqref{eq:45}, leaving the total
dimension of the single particle Hilbert space invariant. In the
thermodynamic limit $L\rightarrow\infty$, the uniform term is
unobservable leaving a point-like Majorana zero mode with ``dimension''
$\delta=1/2$, thus obeying exclusion statistics with parameter
1/2\cite{Haldane1991a,Ye2015,Ye2017}.  Furthermore the \textit{total
  Hilbert space} of $N$ fermion modes has dimension $2^{N}$, but
according to the above calculation the total Hilbert space of $N$
soliton modes has dimension $2^{\delta N}$, so that their quantum
dimension is $2^\delta=\sqrt{2}$, intriguingly coinciding with the
quantum dimension of Majorana vortex mode in a topological
superconductor\cite{Nayak2008,Kitaev2006}.

It should be emphasized that the spatial dependence of $D(x)$ given in
Eq.~\eqref{eq:45} is not at all obvious; this \textit{may be closely
  related to the nonlocal nature of the topological Majorana type
  excitations}, and also distinguishes the Majorana zero mode from other
one-dimensional topological excitations. For example, one can calculate
the dimension density for the complex $Z_2$ soliton models, and it turns
out to be $D(x)-N_D/L=0$ (see the appendix) implying a quantum dimension
$2^0=1$ of the complex $Z_2$ soliton mode. Such a result is also in
agreement of the quantum dimension of abelian anyons\cite{Kitaev2006b}
(note that these solitons are connected to the anyonic excitations in
the fractional quantum Hall effect\cite{lee2007}).  Given all these
coincidences, however, it is still lacking a clear connection between
the "dimension" defined through a renormalized trace of the identity, as
done above, and the quantum dimension.

\section{Conclusion}
\label{sec:conclusion}
As a summary, we consider the soliton mode emerging in one-dimensional
superconductors which are closely related to the fractional
quasiparticles in the two dimensional systems using the idea of edge
soliton as given in Ref.\onlinecite{lee2007}.

For $s$-wave pairing, the phase of the order parameter is twisted by
$\pm \varphi/2$ at the two ends. The mean-field Hamiltonian is analogous
to the Jackiw-Rebbi model, suggesting the existence of fractional spin
excitations. Indeed, by solving the corresponding BdG equation, we find
a single complex $Z_2$ soliton mode carrying a fractional spin number
$c=\varphi/(2\pi)$, although the fermion doubling conceals this
fractional excitation.  Using the exact wavefunctions in the presence of
a complex $Z_2$ soliton, we expand the spin density in a system of
length $L$ with respect to $L^{-1}$ and find that it consists of two
parts, a localized spin density profile with a fractional number $c$ and
a uniform spin density $-c/L$. The uniform distribution disappears in
the thermodynamic limit, but it is essential to keep the spin number
conserved for a finite system. Our analysis is applicable to other
one-dimensional systems with fractional excitations such as the
Jackiw-Rebbi model with particle number conserved, thus provides a
mechanism to solve the puzzle of the emergence of the fractional fermion
number in a finite system with conserved integral quantum number.

For $p$-wave pairing with chemical potential interpolating between the
strong and weak pairing phase, the emergent $Z_2$ soliton mode is of the
Majorana type. By introducing the concept of dimension density of the
single particle Hilbert space, we associate the Majorana zero mode with
1/2 dimension of the \textit{single particle} Hilbert space, which
suggests that it may be understood as an object with 1/2 exclusion
statistics\cite{Haldane1991a,Ye2015,Ye2017}. This result also suggests a
dimension of $\sqrt{2}$ in the total Hilbert space which agrees with the
quantum dimension of the Majorana vortex in a two-dimensional $p$-wave
superconductor. Similar calculations can also be performed for the
complex $Z_2$ soliton, it turns out to be $2^0=1$ coinciding with the
quantum dimension of abelian anyons\cite{Kitaev2006b}.

Finally, strictly speaking the mean-field treatment may not be accurate
for one-dimensional interacting fermions where the Luttinger liquid
theory may be more appropriate. However as shown in
Ref.\onlinecite{Froehlich1988}, the interaction does not interfere with
the appearance of a fractional charge and this suggests that the
argument given in this article can be extended to the framework of
Luttinger liquids.

\acknowledgments{ We would like to thank C. X. Liu, T. Li, G. Su,
  Z. Y. Weng, F. Yang, and Y. Zhou for helpful discussions. F. Ye is
  financially supported by National Nature Science Foundation of China
  11374135, 11774143 and JCYJ20160531190535310.}

\appendix*{}
\section{Derivation of spin density in space}
\label{sec:deriv-spin-dens}
We introduce the following three terms
\begin{align}
\label{eq:46}
\tilde{s}_m(x) = &\frac{\kappa}{2}\sech^2(\kappa x), \nonumber\\
\tilde{s}_-(x) =& \sum_{n}
|\psi^S_{-}(x,\bar{k}^{(-)}_n)|^2, \nonumber\\
\tilde{s}_+(x) =& \sum_n
|\psi^S_{+}(x,\bar{k}^{(+)}_n)|^2,
\end{align}
which correspond to the midgap state, valence band and conduction band
of $\mathcal{H}_1$. They are connected with $s_1(x)$ and $s_2(x)$
defined in Sec.~\ref{sec:fract-spin-distr} as $s_1(x)=\tilde{s}_-(x)$
and $s_2(x)=\tilde{s}_m(x)+\tilde{s}_+(x)$.

Using Eq.~\eqref{eq:20}, we find 
\begin{align*}
|\psi^S_{\pm}(x,k)|^2 =& \frac{\kappa^2\tanh^2(\kappa
  x)+k^2+(\omega+\epsilon)^2}{2L\omega(\omega\mp \epsilon)-2\kappa
  \tanh(\kappa L/2)} \nonumber\\
=& \frac{1}{L} + \frac{(2\kappa/L)-\kappa^2\sech^2(\kappa
   x)}{2L\omega(\omega+\epsilon)-2\kappa} + o(e^{-\kappa L}), 
\end{align*}
then 
\begin{align*}
  \tilde{s}_{\pm}(x) = & \frac{\mathcal{N}_{\pm}}{L} + \sum_n
  \frac{(2\kappa/L)-\kappa^2\sech^2(\kappa
   x)}{2L\omega(\omega\mp \epsilon)-2\kappa} \nonumber\\
=& \frac{\mathcal{N}_{\pm}}{L} + \frac{L}{2\pi}\int d\bar{k}
  \frac{(2\kappa/L)-\kappa^2\sech^2(\kappa
   x)}{2L\omega(\omega\mp \epsilon)-2\kappa} \nonumber\\
&\hspace{3cm}\times\left( 1+ \frac{1}{L} \frac{\partial
                                                           \delta^{(\pm)}(\bar{k})}{\partial\bar{k}} \right) 
\end{align*}
where $\mathcal{N}_{\pm}$ is the number of the allowed momenta in the
conduction and valence bands, respectively. Since the midgap state is
lowered down from the conduction band, $\mathcal{N}_+=2N/L$ and
$\mathcal{N}_-=(2N+1)/L$.

Using the identity
\begin{eqnarray}
\label{eq:47}
\frac{\kappa}{\pi}\int_0^{\infty} dk
\frac{1}{\omega(\omega+\epsilon)} 
= \frac{\varphi}{2\pi}\equiv c, 
\end{eqnarray}
we finally obtain
\begin{align}
\label{eq:48}
\tilde{s}_{+}(x)=& \frac{2N+1-c}{L} -\frac{(1-c)\kappa}{2}\sech^2(\kappa
                   x) + o(L^{-1}), \nonumber\\
\tilde{s}_-(x) =& \frac{2N+1+c}{L} -\frac{c\kappa}{2} \sech^2(\kappa x)
                  + o(L^{-1}).
\end{align}
In the presence of soliton,
$\tilde{s}_-(x)+\tilde{s}_+(x)+\tilde{s}_{m}(x) = \frac{2(2N+1)}{L}$ by
ignoring the higher order terms, which is just $\sum_n|\phi_n(x)|^2$ in
Eqs.~\eqref{eq:28} and \eqref{eq:29} and also the dimension density
$D(x)$ defined for these system(see Eq.~\eqref{eq:44}).

 \bibliographystyle{apsrev}

\end{document}